\documentclass[a4paper,11pt]{article}
\usepackage{pos}
\usepackage{graphicx}
\usepackage{xcolor}
\usepackage{amsmath}
\usepackage{physics}
\usepackage{ragged2e}
\usepackage{caption}
\usepackage{subcaption}
\usepackage[numbers]{natbib}
\usepackage{array}
\newcommand{\nc}{N_\mathrm{c}}

\newcommand{\gev}{\mathrm{GeV}}

\newcommand{\qso}{Q_{\mathrm{s},0}} 
 
\newcommand{\rt}{\mathbf{r}}

\newcommand{\nf}{n_\mathrm{f}}
\newcommand{\seq}{\sum_{q}^{\nf} e_q^2}
\newcommand{\lqcd}{\Lambda_\text{QCD}}

\newcommand{\xij}[1]{\mathbf{x}_{#1}}

\newcommand{\as}{\alpha_\mathrm{s}}

\title{Bayesian constraint of the initial condition for the Balitsky-Kovchegov equation at NLO }

\author*[a,b]{Carlisle Casuga}
\author[c]{Henri H\"anninen}
\author[a,b]{Heikki M\"antysaari}

\affiliation[a]{Department of Physics, University of Jyv\"askyl\"a,  P.O. Box 35, 40014 University of Jyv\"askyl\"a, Finland}

\affiliation[b]{Helsinki Institute of Physics, P.O. Box 64, 00014 University of Helsinki, Finland}

\affiliation[c]{Department of Mathematics and Statistics, University of Jyväskylä,
P.O. Box 35, 40014 University of Jyväskylä, Finland}

\emailAdd{carlisle.doc.casuga@jyu.fi}
\emailAdd{henri.j.hanninen@jyu.fi}
\emailAdd{heikki.mantysaari@jyu.fi}

\abstract{We use Bayesian inference to constrain the parameters describing the initial amplitude input to the Balitsky-Kovchegov evolution equation at next-to-leading order accuracy against precise HERA total inclusive cross section and heavy quark data. The datasets are found to provide stringent constraints and, with consistent NLO treatment, a successful description of the data is obtained. The posterior distributions define the theoretical uncertainites that surround the non-perturbative initial condition and, thus, provide a way to propagate said uncertainties to CGC calculations at NLO. }

\FullConference{XXXII International Workshop on Deep Inelastic Scattering and Related Subjects (DIS2025)\\
24-28 March, 2025\\
Cape Town, South Africa\\}

\begin{document}
\maketitle

\section{Introduction}

Measurements from the upcoming Electron-Ion Collider will access kinematic regions expected to be sensitive to saturation, the non-linear phenomenon of gluon recombination in gluon-dense matter. The high precision expected for these future measurements requires  theoretical modeling and phenomenology in the Color Glass Condensate (CGC) framework to be developed to higher precision. The approach towards this goal includes estimating uncertainties related to the non-perturbative input required by all CGC calculations. While significant progress in promoting reaction-dependent hard factors and evolution equations to the next-to-leading (NLO) accuracy has been made over the past decade, theoretical uncertainties have yet to be provided at NLO. 

In the high-energy limit, deep inelastic scattering (DIS) can be described in the dipole picture, where the dipole-target interaction is expressed as a correlator of Wilson lines. The hard factors that describe the photon splitting into partonic states, including heavy quark states, at NLO in $\alpha_s$ are derived in Refs.~\cite{Beuf:2022ndu,Beuf:2021qqa,Beuf:2021srj}. The energy, or Bjorken-$x$, dependence of the dipole-target scattering amplitude is described by the Balitsky hierarchy of evolution equations. In the mean field limit, the evolved dipoles are provided by the Balitsky-Kovchegov (BK) equation~\cite{Kovchegov:1999yj,Balitsky:1995ub}, which requires a non-perturbative initial input. This initial amplitude has been studied extensively, and many groups have constrained it against HERA data at leading order~\cite{Albacete2009, Albacete2011, Lappi2013}, and NLO with light quarks only~\cite{Beuf2020} and recently with the heavy quarks included~\cite{Hanninen2023}. 

It has been shown in Ref.~\cite{Hanninen2023}  that an excellent description of both the total and heavy quark production cross section can be achieved by a consistent NLO calculation. The full NLO BK equation has been derived in Ref.~\cite{Balitsky:2008zza}.
Furthermore, a subset of higher order corrections enhanced by large transverse logarithms can be resummed, as done e.g. in the kinematically-constrained BK (KCBK) equation~\cite{Beuf:2014uia}. This resummed equation can be derived by imposing time ordering between the consecutive gluon emissions in the evolution. This effectively resums the enhanced double logarithmic terms resulting to an approximative prescription of the NLO BK equation.

Our recent work~\cite{Casuga:2025etc} promotes our leading order analysis \cite{Casuga2024} to partial NLO accuracy using the NLO impact factors combined with the KCBK evolved dipoles. We provide uncertainty estimates for the non-perturbative initial condition of the BK evolution, which is necessary to propagate uncertainties to the computed cross section in all CGC calculations at NLO accuracy.

\section{Setup}


The initial condition for the BK evolution equation can be parametrized using a McLerran-Venugopalan~\cite{Mclerran1994} model inspired functional form as in Ref.~\cite{Lappi2013},
\begin{equation}
        N(\rt, Y=0) = 1-\exp \left[  - \frac{\left(\rt^2 \qso^2\right)^\gamma}{4} \ln \left( \frac{1}{|\rt| \Lambda_\text{QCD}} +  e_c\cdot  e   \right)   \right]. 
\end{equation}
Here $Y$ is the evolution rapidity, the parameter $Q_{s,0}^2$  controls the initial saturation scale, $e_c$ works as an infrared regulator, and $\gamma$ is the anomalous dimension that controls the slope of $N(r)\sim (r^2Q_s^2)^\gamma$ at small dipole size $r$. We focus on the case where $e_c = 1$. The impact parameter dependence of the cross section is replaced by another model parameter, $\int \mathrm{d}^2b \rightarrow\sigma_0/2$, interpreted as a transverse proton area. The running of the strong coupling in the transverse coordinate space is parametrized as 
\begin{equation}
    \as(\xij{ij}^2) = \frac{4 \pi}{ \beta_0 \ln \left[ \left( \frac{\mu_0^2}{\lqcd^2}\right)^{1/c} + \left(\frac{4 C^2}{\xij{ij}^2 \lqcd^2}\right)^{1/c} \right]^c },
    \label{eq:alpha}
\end{equation}
where the $C^2$ is a parameter that controls the speed of evolution of the BK kernel. Here,  $\Lambda_\mathrm{QCD} = 0.241 \ \gev, \beta=(11 \nc - 2N_f)/3$, and, in this work, we use $N_f = 3$. This analysis uses two different schemes in determining the scale, $\xij{ij}^2$, at which the running coupling is evaluated. The setup where the size of the original dipole dictates the scale is the parent dipole running coupling scheme. Meanwhile, one can choose the distance scale to approximatively correspond to the smallest dipole. We call this setup the Bal+SD running coupling scheme, where the Balitsky prescription~\cite{Balitsky:2006wa} defines the BK kernel, see ~\cite{Casuga:2025etc} for details.

This paper presents the probability distributions of the parameters that describe the initial condition obtained by constraining against the most recent combined  deep inelastic $ep$ cross section data~\cite{H12015}, and the charm quark contribution~\cite{H1:2018flt}, from HERA in the region 2.0 $\mathrm{GeV}^{2} < Q^2 < $ 50.0 $ \mathrm{GeV}^2$. These probability distributions are distributions of the posterior as defined by the Bayes' theorem. The posterior can be expressed as a product of the likelihood and the prior. The likelihood function describes the agreement between the constraining dataset and the model while the prior is a function that encodes \textit{prior} knowledge, for example the bounds of the parameter space. 

We use a Bayesian inference setup that is a combination of a Markov Chain Monte Carlo (MCMC) sampler and a Gaussian process emulator (GPE). The GPE is trained to model calculations and is intended to be a faster replacement to the model. The MCMC, with input from the GPE and experimental data, determines the posterior distribution for our model parameters. The posterior distribution is formed from samples in the MCMC chain as it converges to areas of high posterior.  

\section{Results}

\begin{figure}[ht!]
    \centering
    \includegraphics[width=\linewidth]{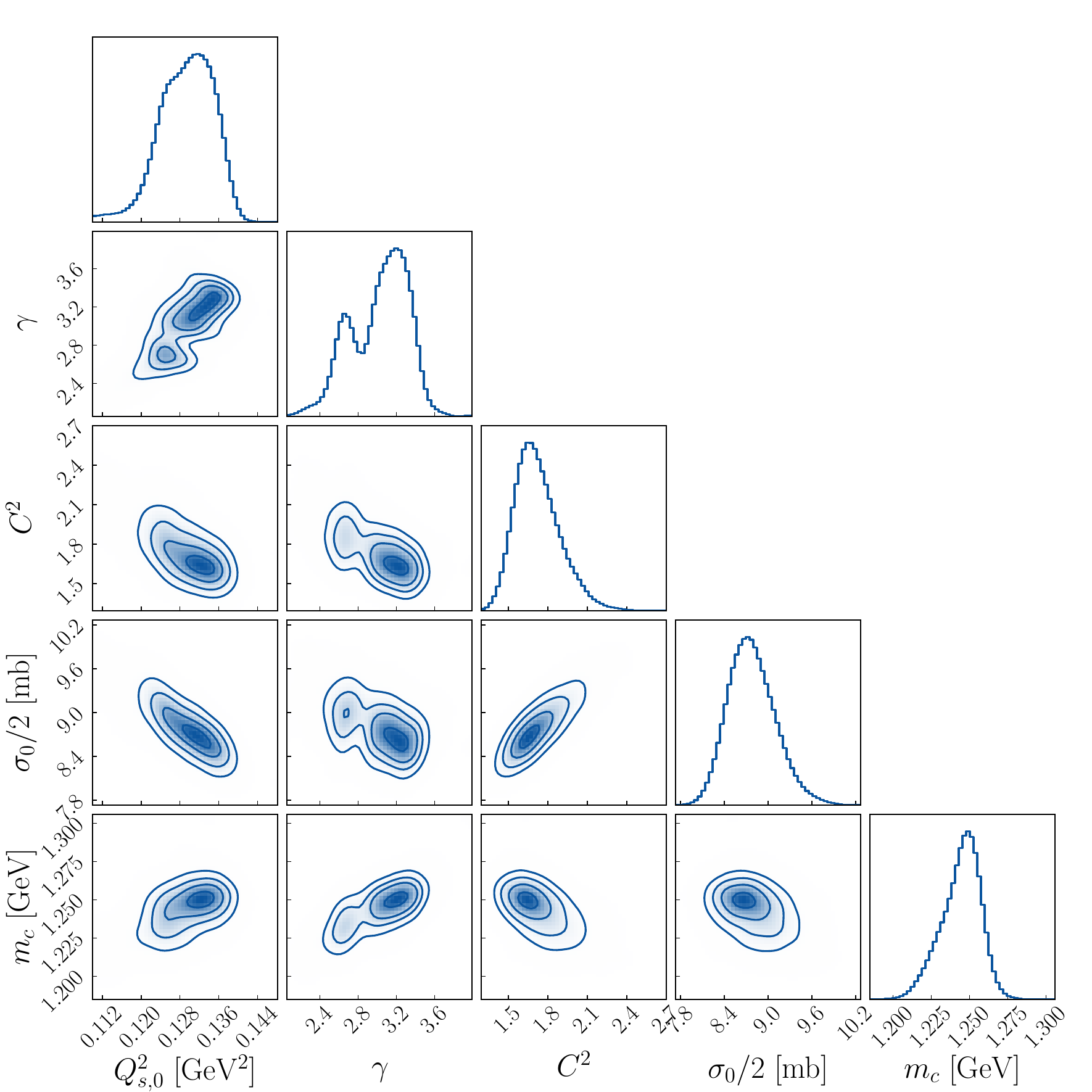}
    \caption{Posterior distribution for the fit using the Bal+SD running coupling scheme.}
    \label{fig:balsd}
\end{figure}

\begin{figure}[ht!]
    \centering
    \includegraphics[width=\linewidth]{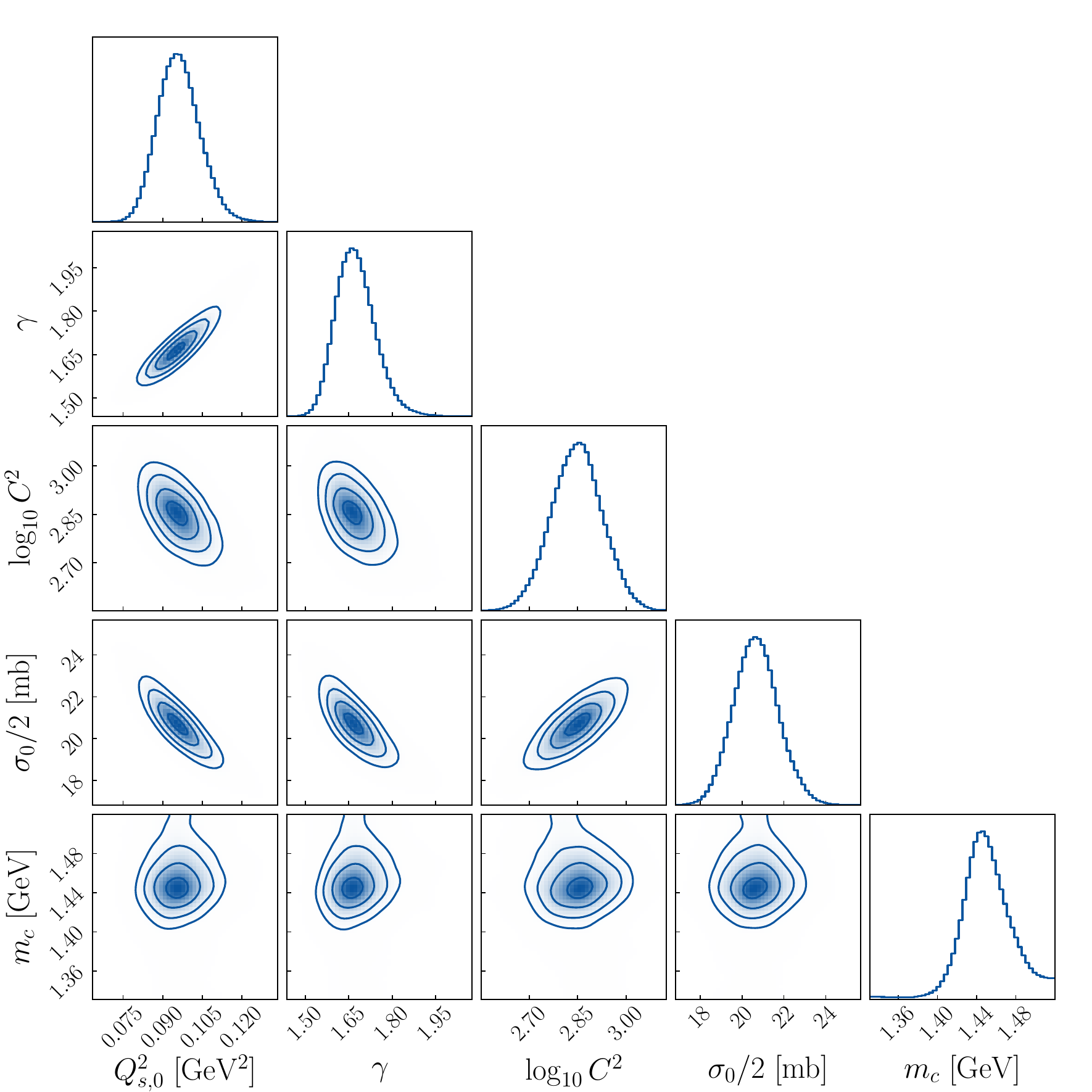}
    \caption{Posterior distribution for the fit using the parent dipole running coupling scheme.}
    \label{fig:pd}
\end{figure}

The total inclusive cross section and charm contribution datasets are found to provide excellent constraints for the initial condition parameters, $\theta = [\qso^2, \gamma, C^2, \sigma_0/2, m_c]$. The charm mass parameter, $m_c$, captures theoretical uncertainties surrounding the fit to charm data. This work features the first successful dipole model fit that constrains to both total and heavy quark data simultaneously. The posterior distributions~\cite{Casuga:2025etc} for the setups using the Bal+SD and the parent dipole running coupling schemes are presented in Figs.~\ref{fig:balsd} and~\ref{fig:pd}, respectively. The corner plot shows 1- and 2- dimensional projections of the 5-dimensional posterior distribution. Maximum-a-posteriori (MAP) values (parameter points with the highest posterior value) along with $\chi^2/\mathrm{dof}$ values are presented in Table~\ref{tab:map}. Furthermore, the fit demonstrates an excellent agreement with the HERA data, where $\chi^2/\mathrm{dof} \sim 1.4$ for the Bal+SD and $\chi^2/\mathrm{dof} \sim 1.2$ for the parent dipole setup. 

\begin{table*}[h!]
    \centering
    \begin{tabular}{|p{7.0cm}|wc{3.0cm}|wc{3.0cm}|}
    \hline
        \textbf{Parameter Description} & \textbf{Bal+SD} & \textbf{parent dipole}  \\ 
    \hline
    \hline
        Initial scale, $\qso^2 \ \mathrm{[GeV}^{2}]$ & 0.124 &  0.090 \\
        
        Anomalous dimension, $\gamma$ & 3.23 &  1.60 \\

        Running coupling scale, $C^{2}$  & 1.74 &  663 \\

        Proton transverse area, $\sigma_{0}/2 \ \mathrm{[mb]}$  & 9.08 &  20.7 \\

        Charm mass, $m_c \ \mathrm{[GeV]}$ &  1.24 &  1.40 \\

    \hline
    Saturation scale, $Q_s^{2}$, at $Y=\ln\left(\frac{1}{0.01}\right)$ $[\mathrm{GeV}^2]$ & 0.196  & 0.199 \\ 

    \hline

    $\chi^2$/d.o.f. values & 1.38 & 1.21 \\
    \hline
    \end{tabular}
    \caption{MAP values for the BK initial condition parametrization obtained using the Balitsky+smallest dipole (labeled as Bal+SD) and parent dipole running coupling schemes. 
    The presented $\chi^2/\mathrm{dof}$ values are obtained when compared to all constraining HERA data, and saturation scales at $Y=\ln \frac{1}{0.01}$ are determined from the definition $N(r^2=2/Q_s^2)=1-e^{-1/2}$. }
    \label{tab:map}
\end{table*}

The Bal+SD scheme, although theoretically-favored, has not previously~\cite{Beuf2020} produced a good description of the HERA data with heavy quarks included. This work is, then, the first successful fit using the Bal+SD scheme at NLO. The evolution speed is controlled by the running of the coupling parametrized according to Eq.~\eqref{eq:alpha}. The Balitsky prescription is expected to slow the evolution speed down, which results to lower $C^2$ values compared to that of the parent dipole scheme. The transverse proton area, $\sigma_0/2$, shown to be negatively correlated with $C^2$, tends to lower values for the parent dipole setup. This sensitivity of the $\sigma_0/2$ and $C^2$ parameters to the running coupling scale is similarly observed in a previous NLO study \cite{Beuf2020} that fits to light quark data.

Our current NLO analysis finds the world data's preference to a steep dipole where $\gamma > 1$, unlike in our previous leading order analysis~\cite{Casuga2024}. While this could produce negative unintegrated gluon distribution values~\cite{Giraud2016}, the BK evolution tames the negativity. 

\section{Summary}

This paper presented the probability distributions of the parameters describing the non-perturbative initial condition to the BK equation at, for the first time, partial NLO accuracy. The parameters were constrained simultaneously to both the total inclusive cross section and charm contribution data from HERA. The posterior distributions extracted for each running coupling setup provide a statistically rigorous way to propagate theoretical uncertainties in NLO CGC calculations. 

A continuation of this work is its promotion to a full NLO analysis, including the previously neglected resummation of large single transverse logarithms and other finite $\alpha_s$ terms. 

\acknowledgments

This work was supported by the Vilho, Yrj\"o, and Kalle V\"ais\"al\"a Foundation, by the Research Council of Finland (Centre of Excellence in Quark Matter, Flagship of Advanced Mathematics for Sensing Imaging and Modelling grant 359208, Centre
of Excellence of Inverse Modelling and Imaging grant
353092, and projects 338263 and 346567), and by the European Research Council (ERC, grant agreements No. ERC-2018-ADG-835105 YoctoLHC, and ERC-2023-101123801 GlueSatLight).
This work used computing resources from
CSC – IT Center for Science in Espoo, Finland and from
the Finnish Grid and Cloud Infrastructure (persistent identifier \texttt{urn:nbn:fi:research-infras-2016072533}). The content of this article does not reflect the official opinion of the European Union and responsibility for the information and views expressed therein lies entirely with the authors.

\bibliographystyle{JHEP}
\bibliography{refs.bib}

\providecommand{\href}[2]{#2}\begingroup\raggedright\begin{thebibliography}{10}

\bibitem{Beuf:2022ndu}
G.~Beuf, T.~Lappi and R.~Paatelainen, \emph{{Massive quarks in NLO dipole factorization for DIS: Transverse photon}}, \href{https://doi.org/10.1103/PhysRevD.106.034013}{\emph{Phys. Rev. D} {\bfseries 106} (2022) 034013} [\href{https://arxiv.org/abs/2204.02486}{{\ttfamily 2204.02486}}].

\bibitem{Beuf:2021qqa}
G.~Beuf, T.~Lappi and R.~Paatelainen, \emph{{Massive quarks in NLO dipole factorization for DIS: Longitudinal photon}}, \href{https://doi.org/10.1103/PhysRevD.104.056032}{\emph{Phys. Rev. D} {\bfseries 104} (2021) 056032} [\href{https://arxiv.org/abs/2103.14549}{{\ttfamily 2103.14549}}].

\bibitem{Beuf:2021srj}
G.~Beuf, T.~Lappi and R.~Paatelainen, \emph{{Massive Quarks at One Loop in the Dipole Picture of Deep Inelastic Scattering}}, \href{https://doi.org/10.1103/PhysRevLett.129.072001}{\emph{Phys. Rev. Lett.} {\bfseries 129} (2022) 072001} [\href{https://arxiv.org/abs/2112.03158}{{\ttfamily 2112.03158}}].

\bibitem{Kovchegov:1999yj}
Y.V.~Kovchegov, \emph{{Small $x$ $F_2$ structure function of a nucleus including multiple pomeron exchanges}}, \href{https://doi.org/10.1103/PhysRevD.60.034008}{\emph{Phys. Rev. D} {\bfseries 60} (1999) 034008} [\href{https://arxiv.org/abs/hep-ph/9901281}{{\ttfamily hep-ph/9901281}}].

\bibitem{Balitsky:1995ub}
I.~Balitsky, \emph{{Operator expansion for high-energy scattering}}, \href{https://doi.org/10.1016/0550-3213(95)00638-9}{\emph{Nucl. Phys. B} {\bfseries 463} (1996) 99} [\href{https://arxiv.org/abs/hep-ph/9509348}{{\ttfamily hep-ph/9509348}}].

\bibitem{Albacete2009}
J.L.~Albacete, N.~Armesto, J.G.~Milhano and C.A.~Salgado, \emph{{Non-linear QCD meets data: A Global analysis of lepton-proton scattering with running coupling BK evolution}}, \href{https://doi.org/10.1103/PhysRevD.80.034031}{\emph{Phys. Rev. D} {\bfseries 80} (2009) 034031} [\href{https://arxiv.org/abs/0902.1112}{{\ttfamily 0902.1112}}].

\bibitem{Albacete2011}
J.L.~Albacete, N.~Armesto, J.G.~Milhano, P.~Quiroga-Arias and C.A.~Salgado, \emph{{AAMQS: A non-linear QCD analysis of new HERA data at small-x including heavy quarks}}, \href{https://doi.org/10.1140/epjc/s10052-011-1705-3}{\emph{Eur. Phys. J. C} {\bfseries 71} (2011) 1705} [\href{https://arxiv.org/abs/1012.4408}{{\ttfamily 1012.4408}}].

\bibitem{Lappi2013}
T.~Lappi and H.~M\"antysaari, \emph{{Single inclusive particle production at high energy from HERA data to proton-nucleus collisions}}, \href{https://doi.org/10.1103/PhysRevD.88.114020}{\emph{Phys. Rev. D} {\bfseries 88} (2013) 114020} [\href{https://arxiv.org/abs/1309.6963}{{\ttfamily 1309.6963}}].

\bibitem{Beuf2020}
G.~Beuf, H.~H\"anninen, T.~Lappi and H.~M\"antysaari, \emph{{Color Glass Condensate at next-to-leading order meets HERA data}}, \href{https://doi.org/10.1103/PhysRevD.102.074028}{\emph{Phys. Rev. D} {\bfseries 102} (2020) 074028} [\href{https://arxiv.org/abs/2007.01645}{{\ttfamily 2007.01645}}].

\bibitem{Hanninen2023}
H.~H\"anninen, H.~M\"antysaari, R.~Paatelainen and J.~Penttala, \emph{{Proton Structure Functions at Next-to-Leading Order in the Dipole Picture with Massive Quarks}}, \href{https://doi.org/10.1103/PhysRevLett.130.192301}{\emph{Phys. Rev. Lett.} {\bfseries 130} (2023) 192301} [\href{https://arxiv.org/abs/2211.03504}{{\ttfamily 2211.03504}}].

\bibitem{Balitsky:2008zza}
I.~Balitsky and G.A.~Chirilli, \emph{{Next-to-leading order evolution of color dipoles}}, \href{https://doi.org/10.1103/PhysRevD.77.014019}{\emph{Phys. Rev. D} {\bfseries 77} (2008) 014019} [\href{https://arxiv.org/abs/0710.4330}{{\ttfamily 0710.4330}}].

\bibitem{Beuf:2014uia}
G.~Beuf, \emph{{Improving the kinematics for low-$x$ QCD evolution equations in coordinate space}}, \href{https://doi.org/10.1103/PhysRevD.89.074039}{\emph{Phys. Rev. D} {\bfseries 89} (2014) 074039} [\href{https://arxiv.org/abs/1401.0313}{{\ttfamily 1401.0313}}].

\bibitem{Casuga:2025etc}
C.~Casuga, H.~H{\"a}nninen and H.~M{\"a}ntysaari, \emph{{Initial condition for the Balitsky-Kovchegov equation at next-to-leading order}}, \href{https://doi.org/10.1103/54zd-hyvg}{\emph{Phys. Rev. D} {\bfseries 112} (2025) 034003} [\href{https://arxiv.org/abs/2506.00487}{{\ttfamily 2506.00487}}].

\bibitem{Casuga2024}
C.~Casuga, M.~Karhunen and H.~Mäntysaari, \emph{Inferring the initial condition for the balitsky-kovchegov equation}, \href{https://doi.org/10.1103/physrevd.109.054018}{\emph{Physical Review D} {\bfseries 109} (2024) } [\href{https://arxiv.org/abs/2311.10491}{{\ttfamily 2311.10491}}].

\bibitem{Mclerran1994}
L.D.~McLerran and R.~Venugopalan, \emph{{Computing quark and gluon distribution functions for very large nuclei}}, \href{https://doi.org/10.1103/PhysRevD.49.2233}{\emph{Phys. Rev. D} {\bfseries 49} (1994) 2233} [\href{https://arxiv.org/abs/hep-ph/9309289}{{\ttfamily hep-ph/9309289}}].

\bibitem{Balitsky:2006wa}
I.~Balitsky, \emph{{Quark contribution to the small-$x$ evolution of color dipole}}, \href{https://doi.org/10.1103/PhysRevD.75.014001}{\emph{Phys. Rev. D} {\bfseries 75} (2007) 014001} [\href{https://arxiv.org/abs/hep-ph/0609105}{{\ttfamily hep-ph/0609105}}].

\bibitem{H12015}
{\scshape H1, ZEUS} collaboration, \emph{{Combination of measurements of inclusive deep inelastic ${e^{\pm }p}$ scattering cross sections and QCD analysis of HERA data}}, \href{https://doi.org/10.1140/epjc/s10052-015-3710-4}{\emph{Eur. Phys. J. C} {\bfseries 75} (2015) 580} [\href{https://arxiv.org/abs/1506.06042}{{\ttfamily 1506.06042}}].

\bibitem{H1:2018flt}
{\scshape H1, ZEUS} collaboration, \emph{{Combination and QCD analysis of charm and beauty production cross-section measurements in deep inelastic $ep$ scattering at HERA}}, \href{https://doi.org/10.1140/epjc/s10052-018-5848-3}{\emph{Eur. Phys. J. C} {\bfseries 78} (2018) 473} [\href{https://arxiv.org/abs/1804.01019}{{\ttfamily 1804.01019}}].

\bibitem{Giraud2016}
B.G.~Giraud and R.~Peschanski, \emph{{Fourier-positivity constraints on QCD dipole models}}, \href{https://doi.org/10.1016/j.physletb.2016.06.033}{\emph{Phys. Lett. B} {\bfseries 760} (2016) 26} [\href{https://arxiv.org/abs/1604.01932}{{\ttfamily 1604.01932}}].

\end{thebibliography}\endgroup

\end{document}